\documentclass[twocolumn]{article}

\usepackage[left=1in,right=1in,top=1in,bottom=1in]{geometry}
\usepackage{tikz}
\usetikzlibrary{calc,patterns,angles,quotes}

\usepackage{algorithm}
\usepackage{algpseudocode}
\usepackage{amsmath}
\usepackage{authblk}
\usepackage{url}
\usepackage{derivative}
\usepackage{booktabs}
\usepackage{graphicx}
\usepackage{caption}
\usepackage{hyperref}
\usepackage{float}
\usepackage{subfigure} 

\usepackage[explicit]{titlesec}
\renewcommand{\thesection}{\Roman{section}}
\renewcommand{\thesubsection}{\thesection.\Alph{subsection}}
\renewcommand{\thesubsubsection}{\thesubsection.\arabic{subsubsection}}

\titleformat{\section}{\large\scshape\centering}{\thesection.\space}{0pt}{#1}[]
\titlespacing*{\section}{0pt}{0.5\baselineskip}{0pt}
\titleformat{\subsection}{\normalsize\itshape}{\Alph{subsection}.\space}{0pt}{#1}[]
\titlespacing*{\subsection}{0pt}{0.5\baselineskip}{0pt}
\titleformat{\subsubsection}{\normalsize\itshape}{\arabic{subsubsection}.\space}{0pt}{#1}[]
\titlespacing*{\subsubsection}{0pt}{0.5\baselineskip}{0pt}
\usepackage{geometry}
\newgeometry{left=2cm, right=2cm, top=2.5cm,bottom=3.5cm}
\makeatletter
\renewcommand{\fnum@figure}{Fig. \thefigure}
\renewcommand{\fnum@table}{Tab. \thetable}
\makeatother

\usepackage{nopageno}
\renewcommand{\abstractname}{}

\title{\textbf{\normalsize Global Optimization for Trajectory Design via Invariant Manifolds in the Earth-Moon Circular Restricted Three-Body Problem}}

\author[1,2]{Flavio Tagliaferri}
\author[1]{Emmanuel Blazquez}
\author[1,3]{Giacomo Acciarini}
\author[1]{Dario Izzo}

\affil[1]{Advanced Concepts Team, European Space Agency, European Space Research and Technology Centre (ESTEC), Keplerlaan 1, 2201 AZ Noordwijk, The Netherlands}
\affil[2]{Sapienza University of Rome, Via Eudossiana 18, 00184, Rome, Italy}
\affil[3]{Surrey Space Centre, University of Surrey, GU2 7XH, Guildford, United Kingdom}
\affil[]{flavio.tagliaferri@pec.it, emmanuel.blazquez@esa.int, g.acciarini@surrey.ac.uk, dario.izzo@esa.int}
\date{}  
\begin{document}

\maketitle

\begin{abstract}
\vspace{-1.3\baselineskip}
\textbf{\emph{\quad Abstract} - 
This study addresses optimal impulsive trajectory design within the Circular Restricted Three-Body Problem (CR3BP), presenting a global optimization-based approach to identify minimum $\Delta V$ transfers between periodic orbits, including heteroclinic connections. By combining a Monotonic Basin Hopping (MBH) algorithm with a sequential quadratic solver in a parallel optimization framework, a wide range of minimum $\Delta V$ transfers are efficiently found. To validate this approach, known connections from the literature are reproduced. Consequently, three-dimensional periodic orbits are explored and a systematic search for minimum propellant trajectories is conducted within a selected interval of Jacobi constants and a maximum time of flight. Analysis of the results reveals the presence of very low $\Delta V$ solutions and showcases the algorithm's effectiveness across various mission scenarios.}
\end{abstract}
\section{Introduction}
Nowadays, as multiple space agencies and private companies are becoming more interested in developing missions far beyond the Earth orbit, trajectory design in multi-body environments becomes a fundamental asset. In particular, the Moon is considered the cornerstone of the near future space exploration \cite{esa_moon_strategy}, with several missions aiming to bring scientific discovery, technology advancement, and acquire the essential skills needed to thrive in a different planetary environment, specifically in preparation for upcoming human missions to other planets, e.g. Mars.
In this context, the Earth-Moon Circular Restricted Three-Body Problem (CR3BP) represents a key mathematical framework for preliminary mission analysis studies, offering families of periodic orbits about the lunar libration points with peculiar operational characteristics, and providing a potential low-energy transport network to move between these orbits, by means of the invariant manifolds \cite{szebehely1967theory}. In the CR3BP, the process of optimal trajectory design can be challenging and traditionally involves the use of Poincaré maps (PM), a tool from dynamical systems theory, which allows the mission designer to visualize the solution space and select promising candidate trajectories. In \cite{koon2000heteroclinic} Koon et al. exploited this tool to compute propellant-free heteroclinic transfers, between $L_1$ and $L_2$ periodic orbits in the planar system.
However, when the full 6-dimensional CR3BP is considered, the use of PMs becomes more demanding due to the higher dimensionality of the system which does not enable a qualitative analysis of the maps \cite{haapala2014representations}. Several optimization strategies were developed to tackle this problem.
For example, in \cite{haapala2016framework} Haapala and Howell implemented a combination of interactive and automated search strategies to locate maneuver-free and low $\Delta V$  transfers between vertical Lyapunov, halo and axial orbits with different Jacobi constant in the Earth-Moon systems, and consequently established a catalog of these transfers. In \cite{li2021hierarchical} Zhaoyu Li, together with the other authors, proposed a global search strategy based on comparing the orbital state of the unstable and stable manifolds, incorporated with low-thrust techniques, to seek a suitable matching point for maneuver application. The obtained velocity increments were then refined to achieve more fuel-optimal transfers using Sequential Quadratic Programming techniques, also deriving the gradient of the constraint to enhance the algorithm convergence accuracy and rapidity. Recently, in \cite{henry2023quasi} D. B. Henry and D.J. Scheeres proposed a robust multiparameter continuation scheme to retrieve the families of transfers between both periodic and quasi-periodic orbits, using the PMs to identify the starting solution.

In this work, we propose a black-box approach based on combining the use of a global optimization algorithm, the Monotonic Basin Hopping (MBH), with the sequential quadratic solver SNOPT \cite{Gill2005} in a parallel optimization framework to perform trajectory design in the CR3BP. More specifically, we focus on impulsive minimum propellant transfers via invariant manifolds, between halo and vertical Lyapunov orbits at the same value of Jacobi constant ($C$), about the $L_1$ and $L_2$ libration points of the Earth-Moon system. To validate the approach, we reproduce trajectories from the literature, both in the planar and spatial scenarios. We consequently perform a systematic search of minimum propellant trajectories in a region of Jacobi constants and accounting for a maximum time of flight. We finally report on the best trajectories found with low propellant consumption, for each transfer type. The proposed approach can be used to retrieve initial low $\Delta V$ trajectories without employing preliminary orbit cartography. 
\section{Theoretical background}
In this section, we begin by describing the dynamical system considered. Subsequently, we delve into the method for retrieving the selected periodic orbits, followed by a discussion on computing stable and unstable manifolds.
\subsection{The Circular restricted three-body problem}
\begin{figure}[tb]
    \centering
    \includegraphics[width=0.8\linewidth]{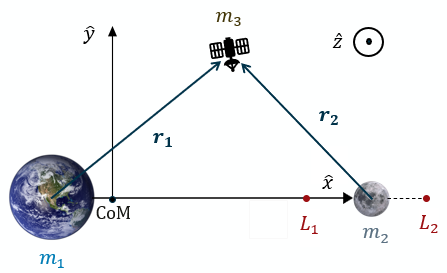}
    \caption{Synodic reference frame and location of the libration points, $L_1$ and $L_2$}
    \label{fig:cr3bp_reference_frame}
\end{figure}

The circular restricted three-body problem describes the motion of an object (satellite) with a negligible mass under the gravitational influence of two other masses $m_1$ and  $m_2$. The larger of these two, $m_1$ (Earth) is called the primary, while $m_2$ (Moon) is called the secondary. The primary and secondary are assumed to orbit their center of mass in a circular orbit. This dynamical system is usually implemented in a non-dimensional coordinate system with distance unit (DU) equal to the distance between $m_1$ and $m_2$ (i.e., $DU=384,400$ km for the Earth-Moon system). The time unit (TU) is such that the mean angular motion of the primary and secondary bodies about the system’s center of mass is unitary ($TU=375,699.79375$ s for the Earth-Moon system). The origin lies at the center of mass of $m_1$ and $m_2$, and the $\hat{x}$ axis points from the primary to the secondary. Aligning parallel to the angular momentum vector of the system is the $\hat{z}$ axis, while the $\hat{y}$ axis completes the right-handed system. This reference frame, known as Synodic reference frame, undergoes a constant rotation around the $\hat{z}$ axis, sharing the same angular velocity as the orbital motion of the primary and secondary, as depicted in Fig.\ref{fig:cr3bp_reference_frame}. Thus, the coordinates of both the primary and secondary remain fixed. The dynamical system can be fully described by \eqref{eq:dynamics}:
\begin{equation}
  \left\{\begin{array}{@{}l@{}@{}}
   \ddot{x}-2\dot{y} = \pdv{U}{x}\\
   \ddot{y}+2\dot{x} =\pdv{U}{y}\\
   \ddot{z} = \pdv{U}{z}\\
  \end{array}\right.\
  \label{eq:dynamics}
\end{equation}
Where:
\begin{equation}
U(x,y,z)=\frac{(x^2+y^2)}{2}+\frac{1-\mu}{r_1}+\frac{\mu}{r_2}
\label{eq:potential}
\end{equation}
represents the potential function, $\mu = \frac{m_2}{m_1+m_2}$ is the mass ratio parameter, which relates the masses of the primary and secondary ($\mu$=0.01215058 for the Earth-Moon system), $r_1$ the distance of the object to the primary and $r_2$ its distance to the secondary.
The CR3BP has five equilibrium points, known as Lagrange points. Three of these points are known as the collinear points and lie on the $\hat{x}$ axis. The two collinear points closest to the moon, $L_1$ and $L_2$ are used for this study. This system admits a first integral of motion, called the Jacobi constant, defined through \eqref{eq:cjac}:
\begin{equation}
C=2U-({\dot{x}}^2+{\dot{y}}^2+{\dot{z}}^2)\ 
\label{eq:cjac}
\end{equation}
This parameter is used in designing heteroclinic connections between different periodic orbits since it is required that the departure and arrival orbits must have the same Jacobi constant. Finally, by constraining the object to move in the xy plane, a simplified framework is retrieved, called Planar Circular Restricted Three-Body Problem (PCR3BP).
\subsection{Periodic orbits}
The computation of periodic orbits is performed using a predictor-corrector method. Initially, the state is numerically integrated, which results in a deviation from periodicity. This initial state is then corrected based on information obtained from the monodromy matrix, defined as the state transition matrix evaluated over one complete period T of the orbit. To achieve this, routines from SEMpy, a Python library specialized in non-Keplerian astrodynamics \cite{blazquez2021sempy}, were utilized and adapted to retrieve the desired values of Jacobi constants for halo, planar, and vertical Lyapunov orbits.
\subsection{Invariant manifolds}
Invariant Manifolds are higher-dimensional dynamical structures that govern the asymptotic flow of motion towards and away from unstable orbits. Those can be computed by exploiting the properties of the monodromy matrix. For each point $\pmb{X_i} = [x,\ y,\ z,\ \dot{x},\dot{y},\dot{z}]$ on the periodic orbit, the initial state of the stable and unstable manifolds, can be computed using \eqref{eq:man_s} and \eqref{eq:man_u}:
\begin{equation}
\pmb{X_i^S}=\pmb{X_i}\pm \epsilon \frac{\Phi_i{,_0} \nu_0^S}{\left\lVert \Phi_i{,_0} \nu_0^S\right\rVert.}
\label{eq:man_s}
\end{equation}
\begin{equation}
\pmb{X_i^U}=\pmb{X_i}\pm \epsilon \frac{\Phi_i{,_0} \nu_0^U}{\left\lVert \Phi_i{,_0} \nu_0^U\right\rVert.}
\label{eq:man_u}
\end{equation}
Where $\Phi_i{,_0}$ is the state transition matrix computed from a reference state $\pmb{X_0}$ up to $\pmb{X_i}$, while  $\nu_0^S$ and $\nu_0^U$ represent the stable ad unstable eigenvectors of the monodromy matrix evaluated at $\pmb{X_0}$, corresponding to the smallest and largest eigenvalue, respectively. The value of the step parameter $\epsilon$ is applied to the whole state in the positive and negative directions, i.e. interior and exterior w.r.t. CR3BP’s center of mass \cite{blazquez2021rendezvous}. In this study, its value is taken constant and equal to 50km for both manifolds, which is a common choice for the Earth-Moon system \cite{gomez1991study}. Finally, the manifolds can be retrieved by numerically integrating the state backward and forward in time for the stable and unstable manifolds, respectively. Intersections of the unstable and stable manifolds result in low-energy transfers.
\section{Trajectory design}
In this section, the proposed approach is explained in detail. First, the chosen problem formulation is addressed, then the specific optimization architecture used is illustrated.
\subsection{Problem formulation}
\begin{figure}[tb]
    \centering
    \includegraphics[width=0.8\linewidth]{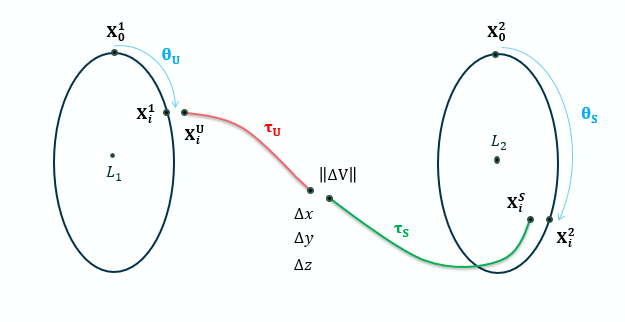}
    \caption{Example of a simplified mission scenario, with all the optimization variables represented}
    \label{fig:mission_scenario}
\end{figure}

In the proposed approach, the search for low $\Delta V$ trajectories between periodic orbits is formulated as a single-objective constrained optimization problem through \eqref{eq:objective}, \eqref{eq:constraints} and \eqref{eq:bounds}:
\begin{align}
\text{minimize} \quad & J=\lVert\Delta V(\textbf{x})\rVert \label{eq:objective} \\
\text{subject to} \quad & \Delta x =0, \Delta y =0, \Delta z =0 \label{eq:constraints} \\
& l_B \le \textbf{x} \le u_B \label{eq:bounds}
\end{align}
where the Euclidean norm of the difference in velocity and the components of the difference in position, i.e. $\Delta x$, $\Delta y$ and $\Delta z$, both computed at the patching point of the unstable and stable manifolds branches, represent the objective function $J$ to be minimized and the three equality constraints to be fulfilled. The decision vector $\textbf{x}$ is composed of four optimization variables: $[\theta_U,\tau_U,\theta_{S,}\tau_S]$. In particular $(\theta_U,\theta_S)$, called the departure and arrival angles, are defined as fractions of the orbit's periods $T_{d_{\text{ep}}}$ and $T_{a_{\text{rr}}}$ through \eqref{eq:theta_u} and \eqref{eq:theta_s}, respectively: 
\begin{equation}
\theta_U =\frac{t_{d{\text{ep}}}}{T_{d{\text{ep}}}}
\label{eq:theta_u}
\end{equation}
\begin{equation}
\theta_S =\frac{t_{a{\text{rr}}}}{T_{a{\text{rr}}}}
\label{eq:theta_s}
\end{equation}
where $t_{d{\text{ep}}}$ and $t_{a{\text{rr}}}$ are the coasting times taken from the reference starting points $\pmb{X_0{^U}}$  and $\pmb{X_0{^S}}$ along the departure and arrival orbits. They are used to reach the states $\pmb{X_i{^U}}$  and $\pmb{X_i{^S}}$  from which the unstable and stable manifolds branches are computed. The coasting times on the unstable and stable manifold branches are represented by $(\tau_U,\tau_S)$. We will refer the sum of $\tau_U$ and $\tau_S$ as the time of flight ($TOF$) of a given trajectory. The optimization variables are bounded between $u_L=[0,0,0,0]$ and $u_B=[1,8,1,8]$. An example of a transfer scenario is depicted in Fig.\ref{fig:mission_scenario}.
\subsection{Optimization architecture}
To solve the presented optimization problem a nested optimization architecture is employed, where a Monotonic Basin Hopping (MBH) algorithm uses sequential quadratic programming (and in particular SNOPT) as its local optimizer. MBH is a stochastic global optimization algorithm \cite{eason1997certain} that works by continuously applying a perturbation vector to the problem's optimization variables in a method called hopping. 
Following each hop, a local optimization is conducted on the new configuration. If this step yields an improved solution, that solution updates the set of optimization variables for the next perturbation. This iterative process is repeated until a maximum number of hops is reached. Here we adopted a modified version of the standard MBH. In particular, the initial decision vector is obtained as the best individual among a population of 200, generated from a random seed. A tolerance of $10^-6$ is set for the three equality constraints. The architecture is then integrated in a parallel optimization framework, where a number of 100 different computational unit, called islands, are used to solve one optimization problem. The best solution is then extracted among the best individual of each island. Note that in principle, multiple trajectories can be retrieved with the solution of a single optimization problem. This could allow the mission designer to make further trade offs based on geometry and time of flight. The resulted optimization architecture as well as the parallel framework was created using the open-source Python library pyGMO \cite{biscani2020parallel}. Moreover, to increase the speed and efficiency of the optimization process, the open-source Python library Heyoka \cite{biscani2021revisiting} was employed for fast integration of ODE systems.
\section{Validation} 
In this section we validate the proposed approach by reproducing established trajectories from the literature. In particular, the algorithm is first tested in the PCR3BP, where heteroclinic connections between planar Lyapunov orbits are sought. Then the CR3BP framework is considered, where heteroclinic connections between periodic orbits become challenging to find. For this reason, a minimum fuel trajectory retrieved through a state-of-the-art method in \cite{henry2023quasi} is set as a reference solution for the validation of the algorithm. Furthermore, each trajectory is characterized by the 
Euclidean norm of the velocity and position differences at the patching point, i.e $\Delta V$ and $\Delta R$, along with the $TOF$. The specific values of the key parameters used in the two validation cases can be found in Tab.~\ref{tab:key_parameters}.
\begin{table}[tb]
\centering
\caption{Key parameters used in the validation cases}
\label{tab:key_parameters}
   \resizebox{\columnwidth}{!}{
   \begin{tabular}{|c|c|c|c|} 
    \hline
    $Case$ & $\mu$ & $\epsilon$ & $TOF [TU]$ \\
    \hline
    PCR3BP & 0.012150585 & $1 \times 10^{-6}$ & - \\
    CR3BP & 0.01215 & $1 \times 10^{-5}$ & $[0,6]$ \\
    \hline
    \end{tabular}
    }
\end{table}
\subsection{Planar trajectories}
\begin{figure}[tb]
    \centering
    \subfigure[]{
        \includegraphics[width=0.46\linewidth]{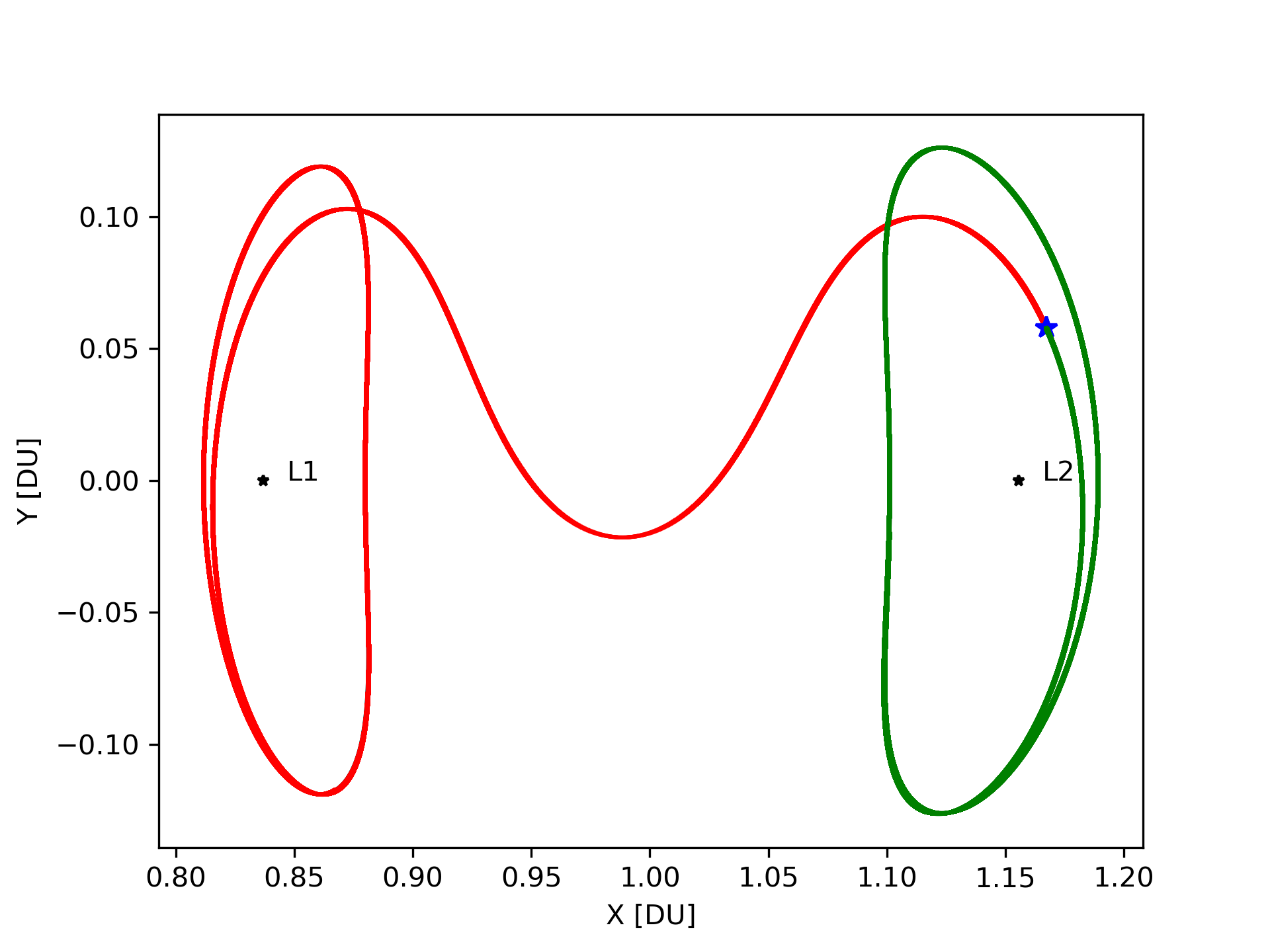}
        \label{fig:case_a}
    }
    \subfigure[]{
        \includegraphics[width=0.46\linewidth]{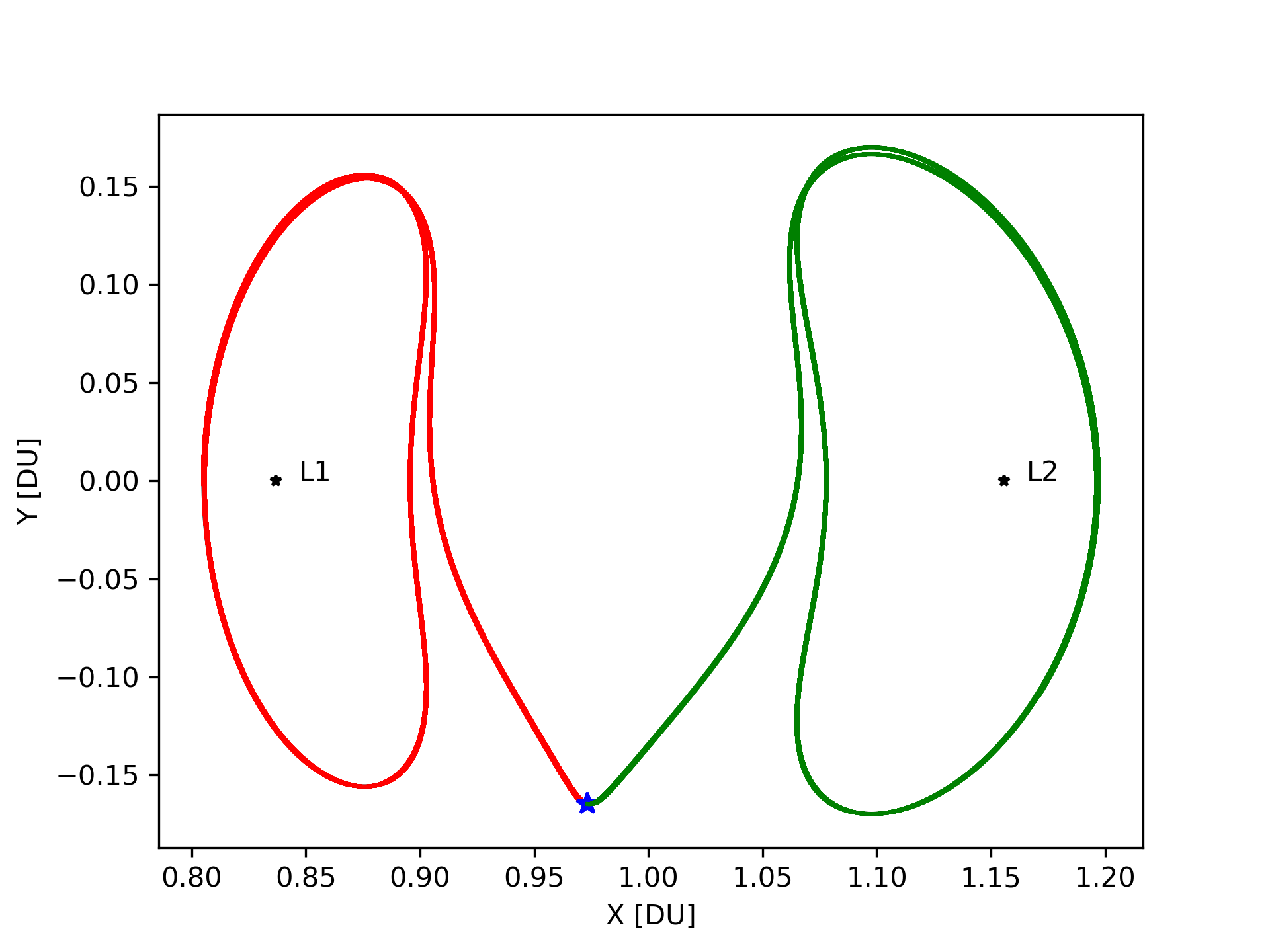}
        \label{fig:case_b}
    }
    \subfigure[]{
        \includegraphics[width=0.55\linewidth]{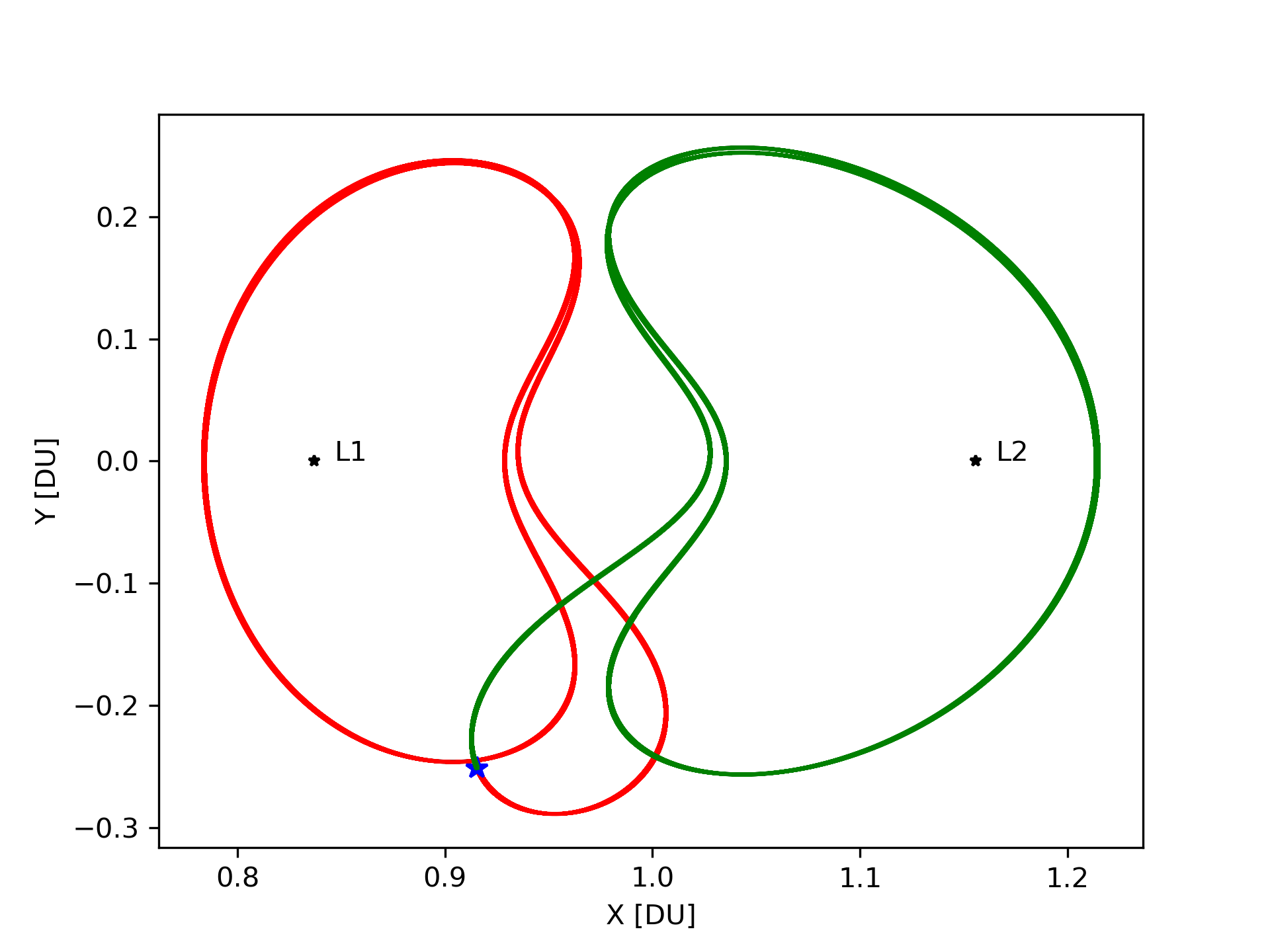}
        \label{fig:case_c}
    }
    \caption{Representation of the obtained numerical heteroclinic connections between planar Lyapunov orbits for cases (a), (b) and (c) in Tab.~\ref{tab:pcr3bp_numerical_results_val}.}
    \label{fig:pcr3bp_val}
\end{figure}

The first task we tackle is the computation of heteroclinic connections between planar Lyapunov orbits, belonging to three different values of the Jacobi constant coming from \cite{barrabes2013numerical}. The detailed numerical results are reported in Tab.~\ref{tab:pcr3bp_numerical_results_val}., while the trajectories are represented in Fig.\ref{fig:pcr3bp_val}., where the unstable and stable branches are depicted in red and green, respectively. As one can see, in all the three cases, the results successfully led to the reconstruction of trajectories with the same qualitative geometry as the ones in the reference paper, up to a precision of about $10^{-4}$ m/s in velocity and of hundreds of meters in position at the patching point.
\begin{table}[tb]
\centering
\caption{Numerical results of the validation in the PCR3BP}
\label{tab:pcr3bp_numerical_results_val}
    \resizebox{\columnwidth}{!}{
    \begin{tabular}{|c|c|c|c|c|} 
    \hline
    $Case$ & $C$ & $\Delta V$ $[m/s]$ & $\Delta R$ $[m]$ & $TOF [dd]$ \\
    \hline
    (a) & 3.130459 & $4.0 \times 10^{-5}$ & 25.53 & 48.075 \\
    (b) & 3.097474 & $1.3 \times 10^{-4}$ & 491.80 & 52.787 \\
    (c) & 3.025554 & $4.3 \times 10^{-4}$ & 360.08 & 75.079 \\
    \hline
    \end{tabular}
    }
\end{table}
\subsection{Spatial trajectories}
\begin{figure}[tb]
    \centering
    \includegraphics[width=0.7\linewidth]{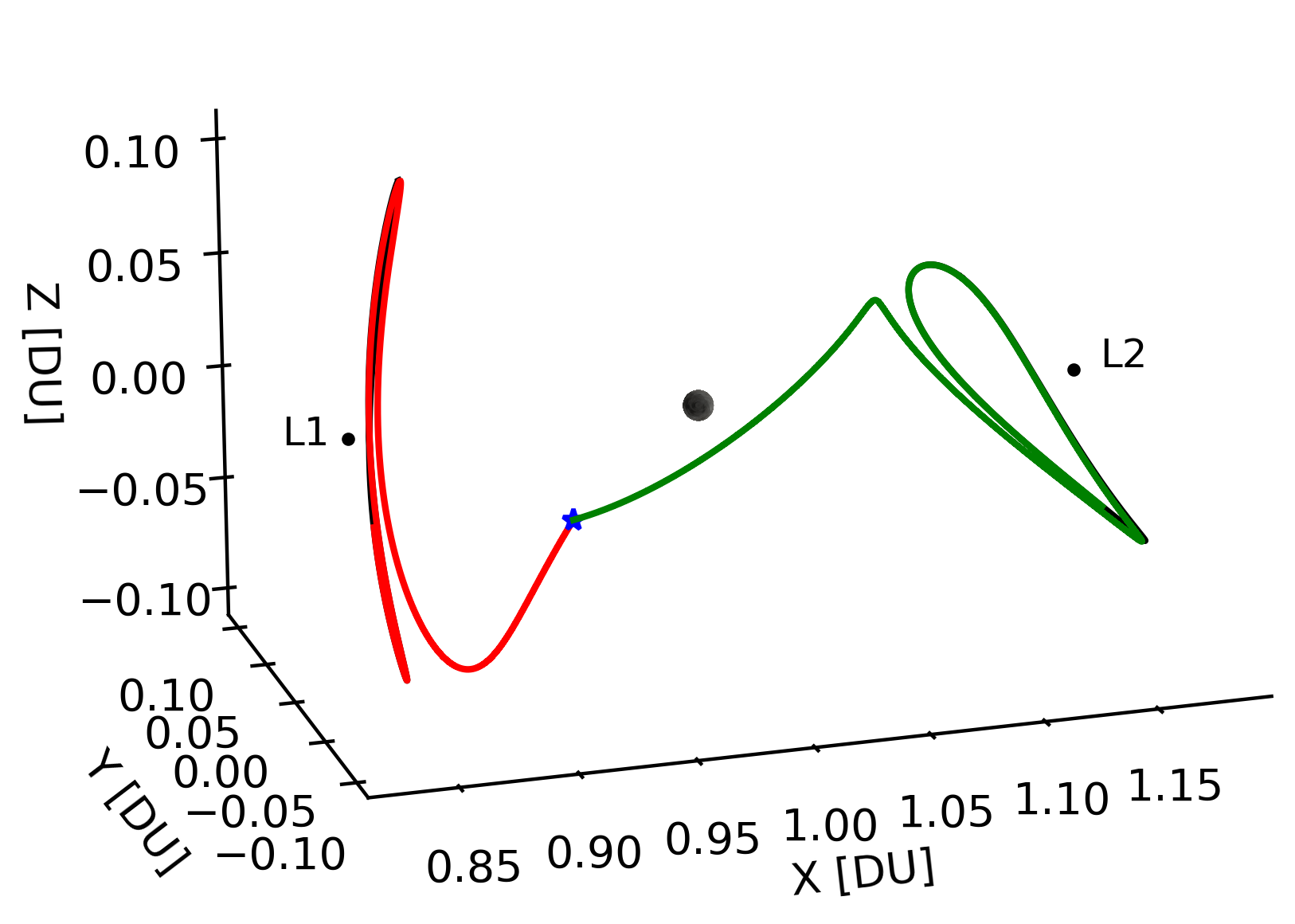}
    \caption{Representation of the trajectory from the reproduced case in Tab.~\ref{tab:cr3bp_numerical_results_val}. from \cite{henry2023quasi}}
    \label{fig:cr3bp_val}
\end{figure}

\begin{figure}[tb]
    \centering
    \subfigure[Three-dimensional trajectory]{
        \includegraphics[width=0.7\linewidth]{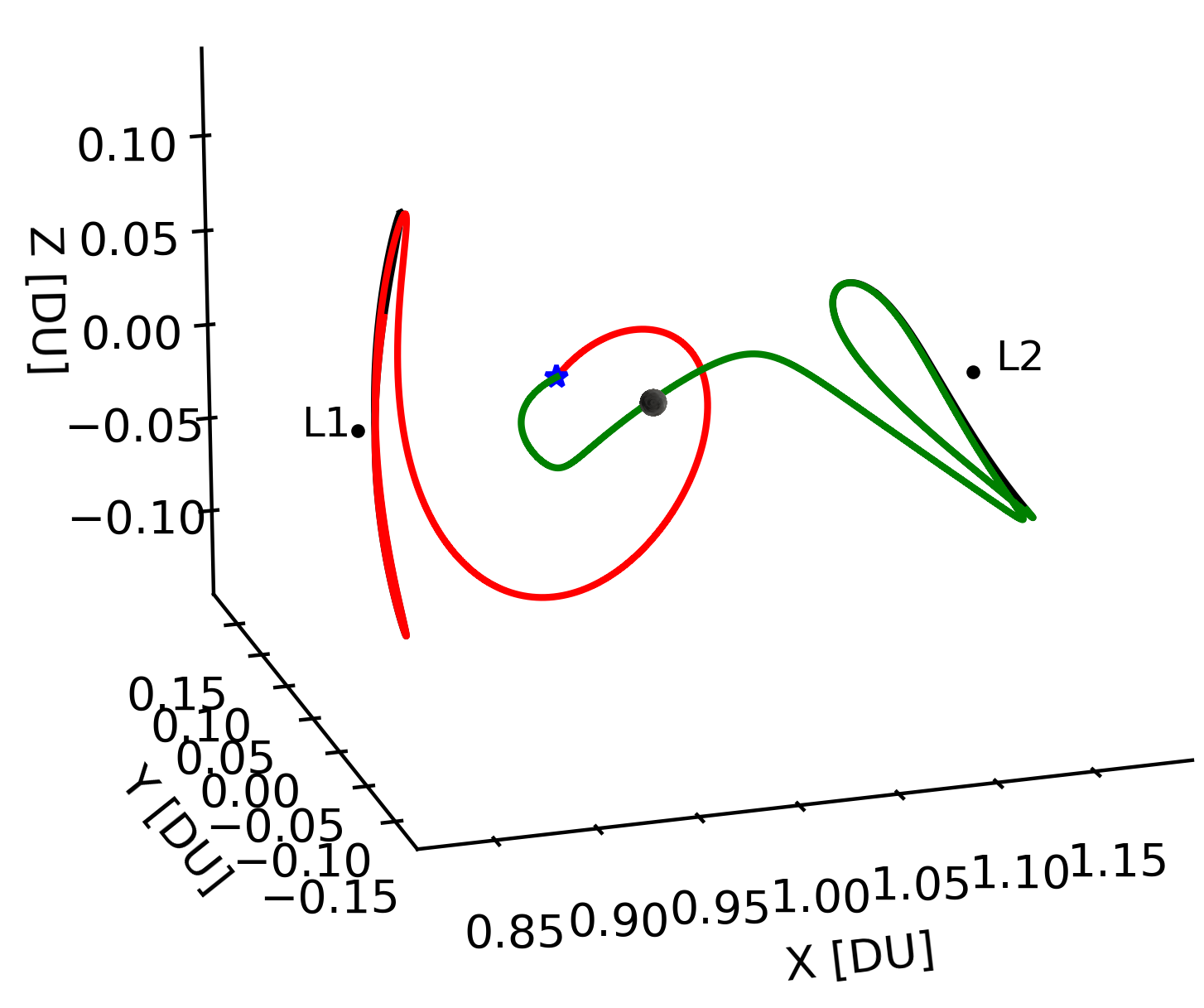}
        \label{fig:best_cr3bp_3dview}
    }
    \subfigure[Projections in the xy and xz planes]{
        \includegraphics[width=0.9\linewidth]{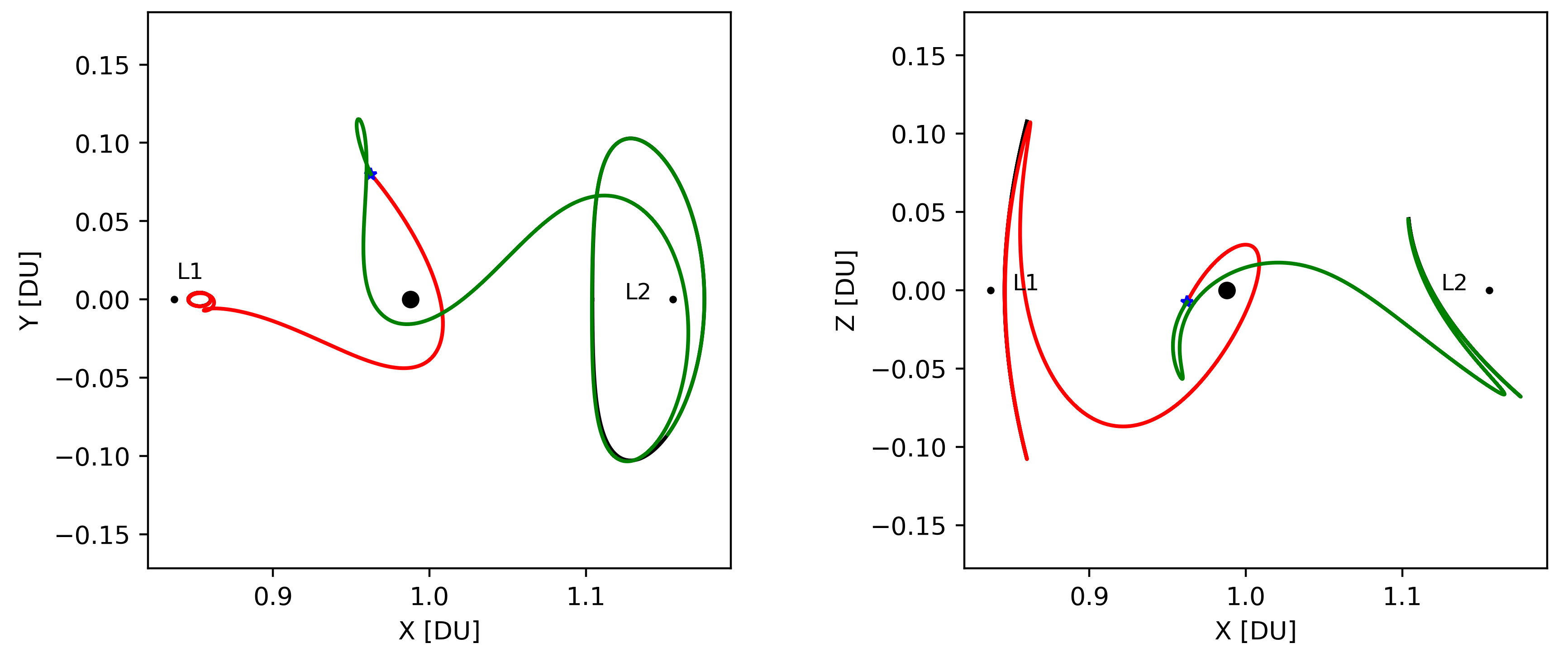}
        \label{fig:best_cr3bp_2dview}
    }
    \caption{Representation of the trajectory from the best min-fuel case in Tab.~\ref{tab:cr3bp_numerical_results_val}., retrieved with the proposed algorithm}
    \label{fig:cr3bp_val_best}
\end{figure}

As a successive test, spatial trajectories are considered. Since finding heteroclinic connections between three-dimensional periodic orbits is very challenging due to the low dimension of their manifolds relative to the phase space, we perform the validation in the CR3BP case by considering the example presented in \cite{henry2023quasi}. This trajectory connects with minimum propellant a $L_1$ vertical Lyapunov and a $L_2$ southern halo orbits at a value of $C$ equal to 3.1328. The results led to the computation of a comparable trajectory in terms of geometry, propellant consumption and time of flight, as showed in Tab.~\ref{tab:cr3bp_numerical_results_val}. Then, the best minimum fuel solution found by our approach is also retrieved. As one can see from Tab.~\ref{tab:cr3bp_numerical_results_val}., the best minimum fuel solution found by the presented algorithm requires about 45\% less fuel and about 13\% longer $TOF$. The corresponding trajectories are depicted in Fig.\ref{fig:cr3bp_val}. and in Fig.\ref{fig:cr3bp_val_best}.
\begin{table}[tb]
\centering
\caption{Numerical results of the validation in the CR3BP}
\label{tab:cr3bp_numerical_results_val}
    \resizebox{\columnwidth}{!}{
    \begin{tabular}{|c|c|c|c|c|}
    \hline
    $Case$ & $C$ & $\Delta V$ [m/s] & $\Delta R$ [m] & $TOF[dd]$ \\
    \hline
    Reference \cite{henry2023quasi} & 3.1328 & 280.47 & - & 39.570 \\
    Reproduced & 3.1328 & 280.70 & 14.38 & 39.564 \\
    Best min-fuel & 3.1328 & 149.10 & 64.04 & 44.789 \\
    \hline
    \end{tabular}
     }
\end{table}
\section{Applications}
In this section, we apply the algorithm to three different transfer types: halo to halo, vertical Lyapunov to vertical Lyapunov and halo to vertical Lyapunov. Our specific goal is to identify the global optimal trajectories that the method can produce, for each transfer type. To achieve this, we systematically explore minimum fuel trajectories between halo and vertical Lyapunov orbits. This investigation covers a Jacobi constant range of $[3.057, 3.152]$ and a maximum time of flight of approximately $70$ days. Notably, both the interior and exterior manifolds are considered in this study. The upper bound of the Jacobi constant corresponds to the value of the bifurcation of the $L_2$ halo family from the Lyapunov family \cite{haapala2016framework}. A step of $6 \times 10^{-4}$ is used to determine the orbit's Jacobi constant and it comes from a trade-off between the number of trajectories obtained and the dimension of successive orbits, i.e. the z-amplitudes. Each mission scenario is identified by:
\begin{enumerate}
    \item The departure and arrival orbits characterized by a value of $C$ and their respective libration points;
    \item The Family of each orbit;
    \item The types of manifolds branches used.
\end{enumerate}
Based on these characteristics, a compact notation is introduced in Tab.~\ref{tab:mission_scenarios_notation}. to identify a particular mission scenario. 
\begin{table}[H]
    \centering
    \small
    \caption{Description of the notation adopted to identify a mission scenario}
    \label{tab:mission_scenarios_notation}
    \begin{tabular}{|c|c|} 
    \hline
    $Notation$ & $Description$\\
    \hline
    NH1 & $L_1$ northern halo\\
    SH1 & $L_1$ southern halo\\
    V1 & $L_1$ vertical Lyapunov\\
    NH2 & $L_2$ northern halo\\
    SH2 & $L_2$ southern halo\\
    V2 & $L_2$ vertical Lyapunov\\
    + & interior manifold\\
    - & exterior manifold\\
    outward & $L_1$ to $L_2$\\
    return & $L_2$ to $L_1$\\
    \hline
    \end{tabular}
\end{table}
For example, with $[NH1-,V2+]$ we describe a trajectory obtained departing from a $L_1$ northern halo orbit moving on the exterior unstable manifold branch and arriving to a $L_2$ vertical Lyapunov orbit using the interior stable manifold branch. Trajectories which either exhibit a $\Delta R \geq 1$ km or impact on the Moon surface are discarded and excluded from the analysis. In the following subsections, we report the global optimal trajectories discovered with this algorithm, for each transfer type.
\subsection{Halo to halo trajectories}
\begin{figure}[tb]
    \centering
    \includegraphics[width=1\linewidth]{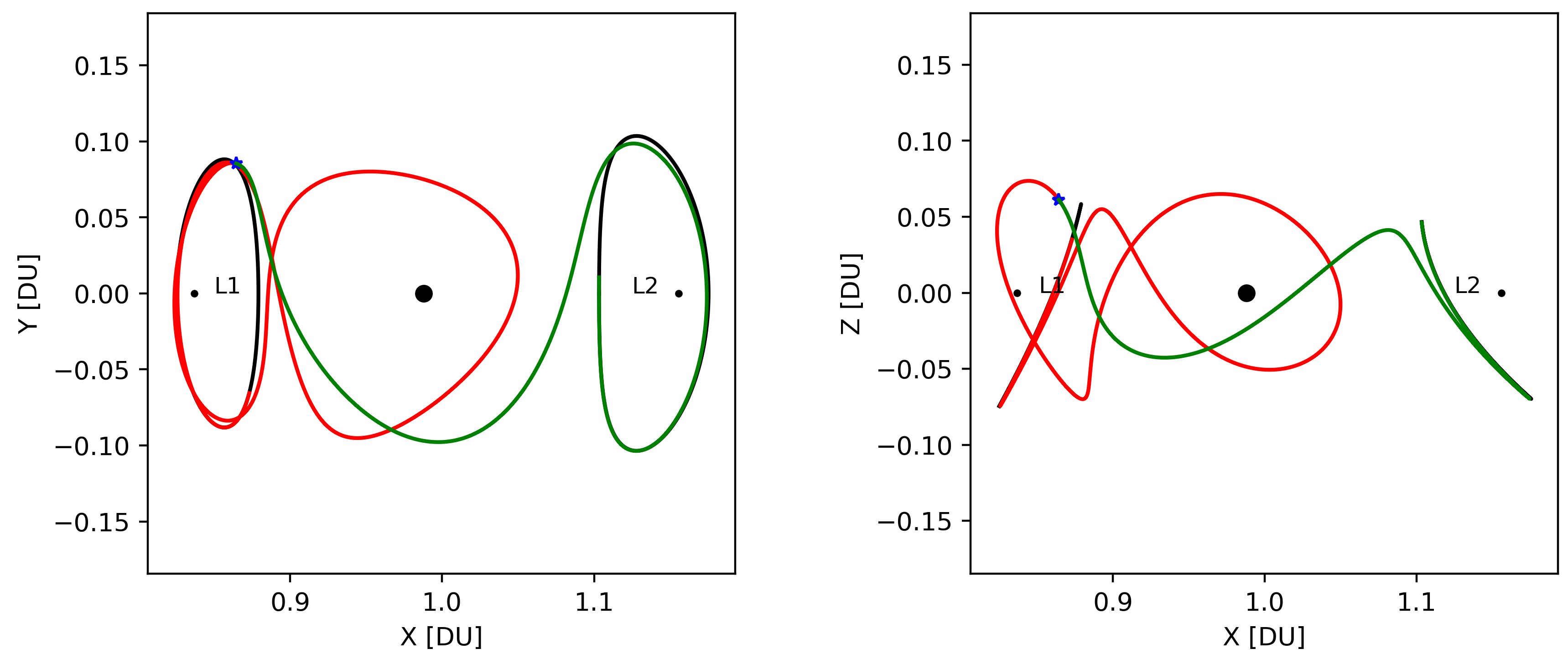}
    \caption{Representation of the global optimum solution for outward trajectories, corresponding to the mission scenario $[SH1-, SH2-]$ in Tab.~\ref{tab:hh_trajectory_results_out}.}
    \label{fig:hh_best_out}
\end{figure}

\begin{figure}[tb]
    \centering
    \includegraphics[width=1\linewidth]{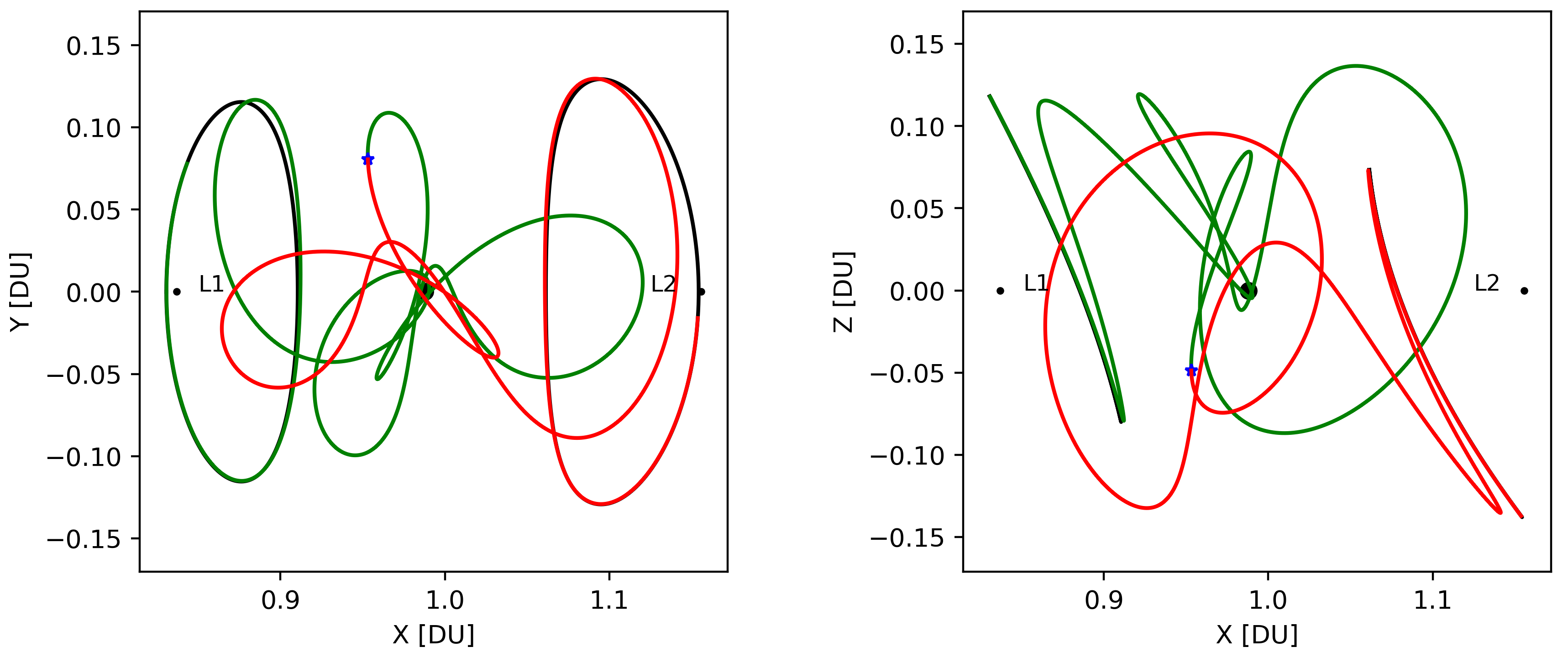}
    \caption{Representation of the global optimum solution for return trajectories, corresponding to the mission scenario $[SH2-, NH1+]$ in Tab.~\ref{tab:hh_trajectory_results_out}.}
    \label{fig:hh_best_ret}
\end{figure}

The best solutions in terms of velocity increment for outward and return trajectories are reported in Tab.~\ref{tab:hh_trajectory_results_out}. Observing the results, we found several trajectories with remarkably low fuel requirements. Among them, the trajectories with the lowest $\Delta V$ are then represented in Fig.\ref{fig:hh_best_out}. and Fig.\ref{fig:hh_best_ret}. Furthermore, the algorithm exhibits a convergence toward outcomes that align with those achievable through the application of CR3BP's symmetries \cite{miele2010revisit}. For example, the successive application of the image of the xy-plane and backward image of the xz-plane to the $[NH1+, SH2-]$ would produce the $[NH2-, SH1+]$ trajectory, while the use of only the backward image of the xz-plane would result in the $[SH2-, NH1+]$, with the same $C$, $\Delta V$ and $TOF$ values. Consistent findings emerged from solving the optimization problems corresponding to these three mission scenarios. This observed behavior validates the obtained results and could offer an opportunity to streamline the investigation of mission scenarios when employing this numerical method.
\begin{table}[tb]
    \centering
    \caption{Numerical results for the optimal halo to halo trajectories, in outward and return mission scenarios}
    \label{tab:hh_trajectory_results_out}
    \resizebox{\columnwidth}{!}{
    \begin{tabular}{|c|c|c|c|c|}
    \hline
    \textbf{Mission scenario} & \textbf{$C$} & $\boldsymbol{\Delta V}$ [m/s] & $\boldsymbol{\Delta R}$ [m] & \textbf{$TOF$ [dd]} \\
    \hline
    $[NH1-,NH2-]$ & 3.132397 & 3.000 & 110.76 & 50.976 \\
    $[NH1+,SH2-]$ & 3.080797 & 2.980 & 9.02 & 64.410 \\
    $[SH1-,SH2-]$ & 3.131797 & 2.684 & 144.10 & 51.077 \\
    $[SH1+,NH2-]$ & 3.071197 & 4.991 & 0.06 & 61.276 \\
    $[NH2+,NH1-]$ & 3.129397 & 3.903 & 446.72 & 49.83 \\
    $[SH2-,NH1+]$ & 3.080797 & 2.980 & 44.49 & 64.410 \\
    $[SH2+,SH1+]$ & 3.132397 & 2.999 & 414.86 & 50.976 \\
    $[NH2-,SH1+]$ & 3.080797 & 5.029 & 31.70 & 64.405 \\
    \hline
    \end{tabular}%
    }
\end{table}
\subsection{Vertical Lyapunov to vertical Lyapunov trajectories}
\begin{figure}[tb]
    \centering
    \includegraphics[width=1\linewidth]{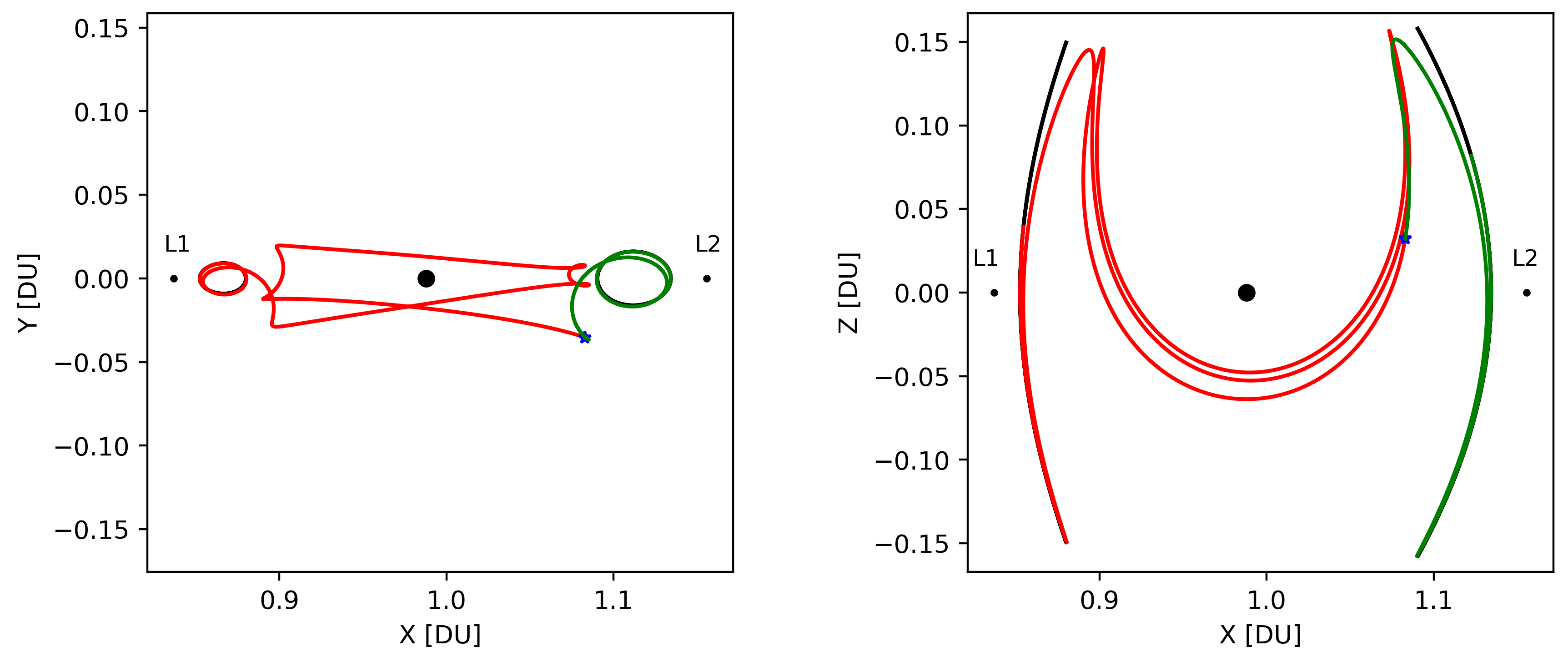}
    \caption{Representation of the global optimum solution for outward trajectories, corresponding to the mission scenario $[V1-,V2-]$ in Tab.~\ref{tab:vv_trajectory_results_out}.}
    \label{fig:vv_best_out}
\end{figure}

\begin{figure}[tb]
    \centering
    \includegraphics[width=1\linewidth]{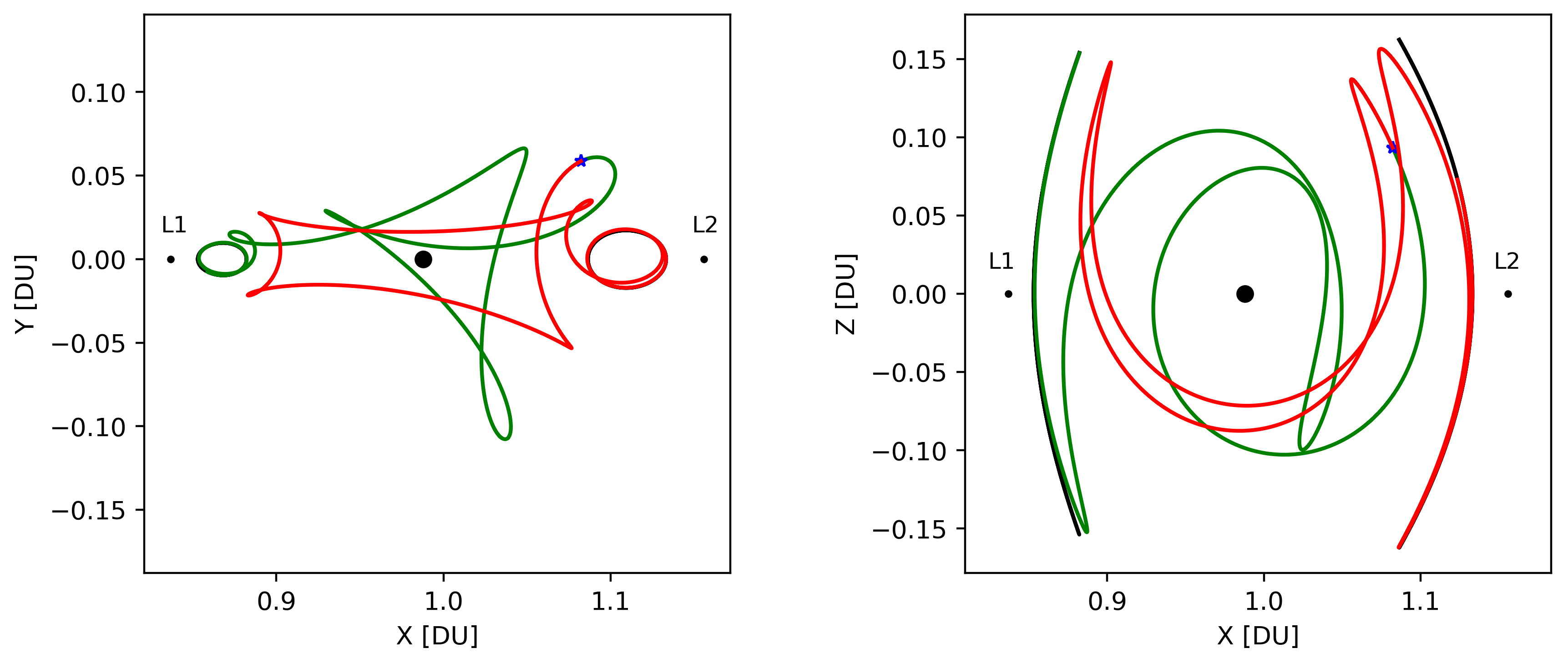}
    \caption{Representation of the global optimum solution for return trajectories, corresponding to the mission scenario $[V2+,V1+]$ in Tab.~\ref{tab:vv_trajectory_results_out}.}
    \label{fig:vv_best_ret}
\end{figure}

The second application we addressed involves computing trajectories connecting vertical Lyapunov orbits. The best solutions in terms of velocity increment for the outward and return scenarios are reported in Tab.~\ref{tab:vv_trajectory_results_out}. Among them, the trajectories with the lowest $\Delta V$ are then represented in Fig.\ref{fig:vv_best_out}. and Fig.\ref{fig:vv_best_ret}. In particular, the mission scenario $[V2+,V1+]$ shows a $\Delta V$ of 4.128 m/s which is, to the best of our knowledge and based of the literature we read, a value remarkably low for this type of transfer. As for halo to halo trajectories, the algorithm demonstrates adherence to the symmetry properties of the CR3BP.
\begin{table}[tb]
    \centering
    \caption{Numerical results for the optimal vertical to vertical Lyapunov trajectories, in outward mission scenarios}
    \label{tab:vv_trajectory_results_out}
    \resizebox{\columnwidth}{!}{
    \begin{tabular}{|c|c|c|c|c|}
    \hline
    \textbf{Mission scenario} & \textbf{$C$} & $\boldsymbol{\Delta V}$ [m/s] & $\boldsymbol{\Delta R}$ [m] & \textbf{$TOF$ [dd]} \\
    \hline
    $[V1+,V2+]$ & 3.109600 & 39.629 & 95.48 & 41.583 \\
    $[V1-,V2+]$ & 3.086799 & 13.068 & 35.23 & 23.253 \\
    $[V1+,V2-]$ & 3.082600 & 9.150 & 1.16 & 38.803 \\
    $[V1-,V2-]$ & 3.089200 & 6.961 & 430.75 & 49.406 \\
    $[V2+,V1+]$ & 3.084400 & 4.128 & 0.50 & 63.482 \\
    $[V2-,V1+]$ & 3.104800 & 61.037 & 3.38 & 26.467 \\
    $[V2+,V1-]$ & 3.089200 & 6.962 & 2.27 & 49.406 \\
    $[V2-,V1-]$ & 3.086799 & 13.0776 & 25.45 & 23.251 \\
    \hline
    \end{tabular}%
    }
\end{table}
\subsection{Halo to vertical Lyapunov trajectories} 
\begin{figure}[tb]
    \centering
    \includegraphics[width=1\linewidth]{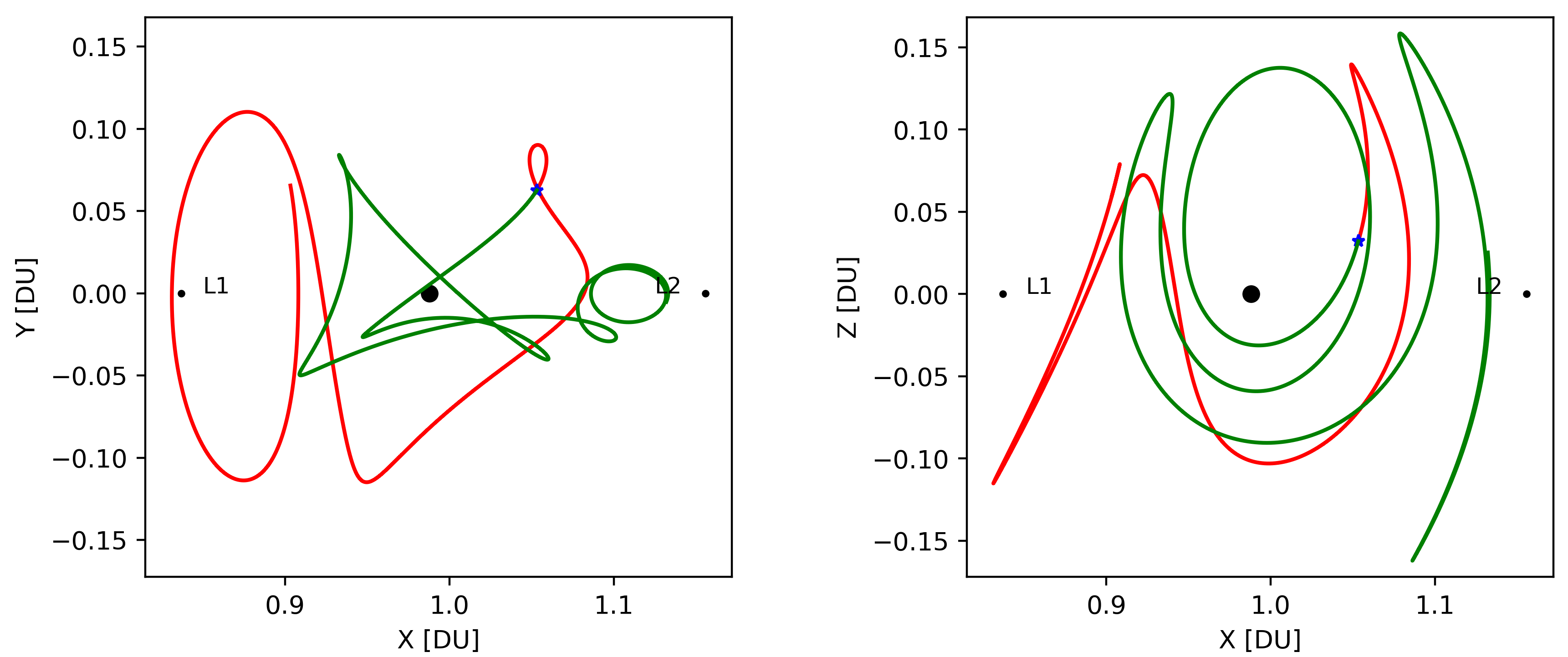}
    \caption{Representation of the global optimum solution for outward trajectories, corresponding to the mission scenario $[SH1+,V2-]$ in Tab.~\ref{tab:hv_trajectory_results_out}.}
    \label{fig:hv_best_out}
\end{figure}

\begin{figure}[tb]
    \centering
    \includegraphics[width=1\linewidth]{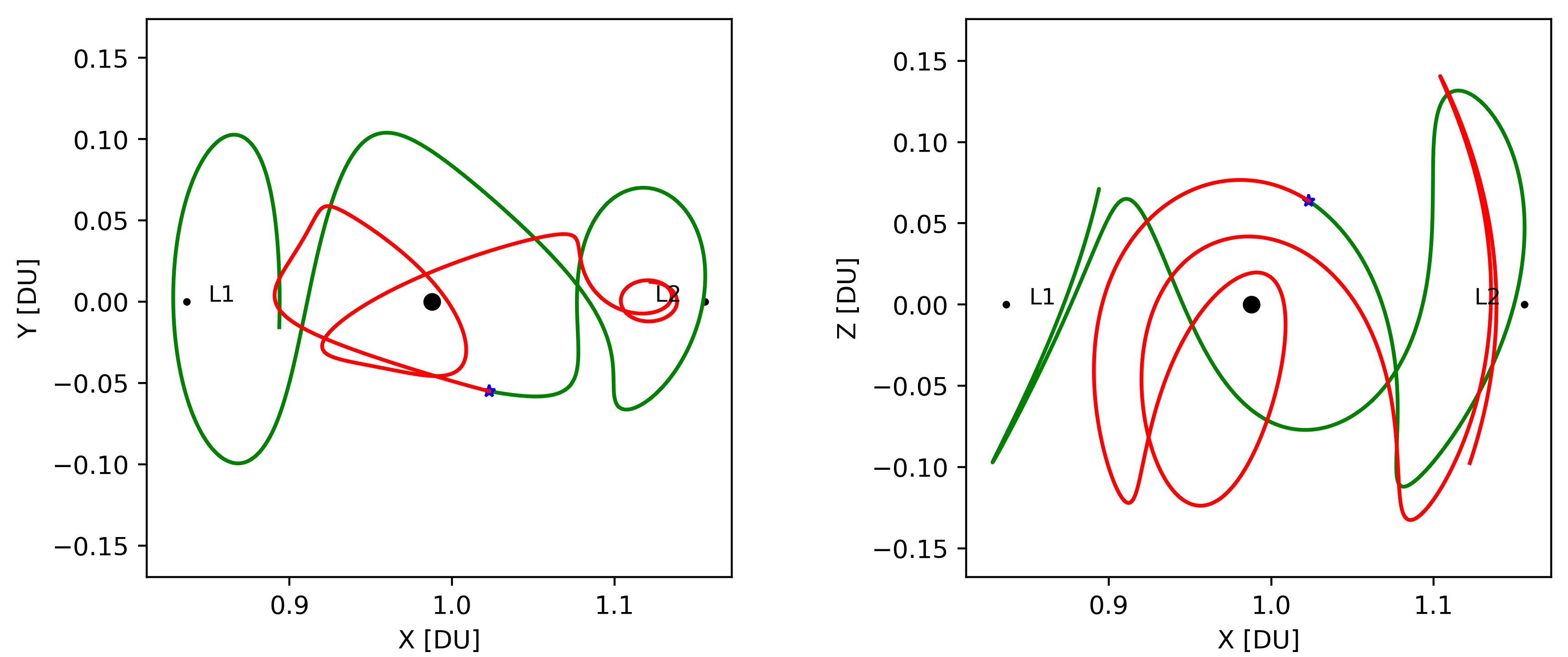}
    \caption{Representation of the global optimum solution for return trajectories, corresponding to the mission scenario $[V2+,SH1+]$ in Tab.~\ref{tab:hv_trajectory_results_out}.}
    \label{fig:hv_best_ret}
\end{figure}

The last application concerns transfers associated with halo orbits and vertical Lyapunov orbits. The best solutions in terms of velocity increment for the outward and return scenarios are reported in Tab.~\ref{tab:hv_trajectory_results_out}. Among them, the trajectories with the lowest $\Delta V$ are then represented in Fig.\ref{fig:hv_best_out}. and Fig.\ref{fig:hv_best_ret}. Higher values of the $\Delta V$ are noticed compared to previous transfer types. Also in this scenario, the algorithm follows symmetry properties, as one can see in Tab.~\ref{tab:hv_trajectory_results_out}.
\begin{table}[htb!]
    \centering
    \caption{Numerical results for the optimal halo to vertical Lyapunov trajectories, in outward mission scenarios}
    \label{tab:hv_trajectory_results_out}
    \resizebox{\columnwidth}{!}{
    \begin{tabular}{|c|c|c|c|c|}
    \hline
    \textbf{Mission scenario} & \textbf{$C$} & $\boldsymbol{\Delta V}$ [m/s] & $\boldsymbol{\Delta R}$ [m] & \textbf{TOF [dd]} \\
    \hline
    $[NH1+,V2-]$ & 3.084997 & 29.338 & 6.37 & 58.522 \\
    $[NH1-,V2-]$ & 3.138997 & 37.159 & 269.83 & 63.321 \\
    $[SH1+,V2-]$ & 3.084397 & 29.283 & 0.11 & 58.630 \\
    $[SH1-,V2-]$ & 3.108397 & 32.678 & 187.17 & 63.670 \\
    $[V2+,NH1+]$ & 3.083197 & 31.950 & 13.89 & 58.866 \\
    $[V2+,NH1-]$ & 3.086197 & 30.731 & 38.40 & 58.317 \\
    $[V2+,SH1+]$ & 3.106597 & 20.939 & 49.32 & 64.209 \\
    $[V2+,SH1-]$ & 3.086197 & 30.731 & 0.14 & 58.317 \\
    \hline
    \end{tabular}%
    }
\end{table}
\section{Conclusions}
In this study, we introduce a general black-box methodology for trajectory design within the framework of the CR3BP. Our validation demonstrates that the proposed approach yields low-cost trajectories that are comparable to those found in the literature. Leveraging a direct problem formulation and an optimized computational architecture, we efficiently compute a large number of trajectories across various scenarios in an automated manner.
The numerical results reveal very low $ \Delta V$ solutions, but also alternative trajectories that could serve as initial approximations for existing methodologies. Remarkably, despite its reliance on a numerical approach, the algorithm demonstrates convergence toward the symmetries of the CR3BP, achieving a certain precision. Furthermore, our proposed approach suggests an exploration of a broader search space compared to traditional methods that rely on Poincaré maps. Indeed, the algorithm we employed operates without the need for defining a surface of section or imposing problem-dependent constraints to reduce system dimensionality. These unique characteristics could facilitate the uncovering of new fuel-optimal solutions, as evidenced by the numerical results for trajectories from vertical Lyapunov to southern halo orbits. Our future research will delve into a more detailed study and analysis of this property, along with an investigation into the convergence limits of the algorithm.

\bibliographystyle{ISSFD_v01}
\bibliography{references}
\end{document}